\documentclass[conference]{IEEEtran}
\IEEEoverridecommandlockouts
\usepackage{cite}
\usepackage{amsmath,amssymb,amsfonts}
\usepackage{algorithmic}
\usepackage{graphicx}
\usepackage{textcomp}
\usepackage{xcolor}

\usepackage{hyperref}
\usepackage{enumitem}
\usepackage{tabularx}
\usepackage{textcomp}
\usepackage{tcolorbox}
\usepackage{amsmath}
\usepackage{amsfonts}
\usepackage{booktabs}
\usepackage{pgf-pie}
\usepackage{multirow}

\usepackage{balance}
\def\BibTeX{{\rm B\kern-.05em{\sc i\kern-.025em b}\kern-.08em
    T\kern-.1667em\lower.7ex\hbox{E}\kern-.125emX}}
\begin{document}

\newcommand{\zq}[1]{\textcolor{cyan}{[Zhiqing: #1]}}
\newcommand{\slhe}[1]{\textcolor{red}{[Shilin: #1]}}
\newcommand{\hw}[1]{\textcolor{yellow}{[hw: #1]}}
\newcommand{\zqtext}[1]{\textcolor{cyan}{#1}}
\newcommand{\hx}[1]{\textcolor{green}{[Haoxuan: #1]}}
\title{An Empirical Study on Package-Level Deprecation in Python Ecosystem}

\author{
\IEEEauthorblockN{Zhiqing Zhong\IEEEauthorrefmark{2}, Shilin He\IEEEauthorrefmark{5}\IEEEauthorrefmark{1}, Haoxuan Wang\IEEEauthorrefmark{2}, Boxi Yu\IEEEauthorrefmark{2}, Haowen Yang\IEEEauthorrefmark{2}, 
Pinjia He\IEEEauthorrefmark{2}\IEEEauthorrefmark{1}\thanks{\hspace{-2ex}\IEEEauthorrefmark{1}Pinjia He and Shilin He are the corresponding authors.}\vspace{2ex}}

\IEEEauthorblockA{
    \IEEEauthorrefmark{2}School of Data Science, The Chinese University of Hong Kong, Shenzhen (CUHK-Shenzhen), China\\
    \IEEEauthorrefmark{5}Microsoft, Beijing, China\\
    \{zhiqingzhong, haoxuanwang, boxiyu, 222010523\}@link.cuhk.edu.cn\\~~shilhe@microsoft.com, ~~hepinjia@cuhk.edu.cn
}
}

\maketitle

\begin{abstract}
Open-source software (OSS) plays a crucial role in modern software development. Utilizing OSS code can greatly accelerate software development, reduce redundancy, and enhance reliability. Python, a widely adopted programming language, is renowned for its extensive and diverse third-party package ecosystem. However, a significant number of OSS packages within the Python ecosystem are in poor maintenance, leading to potential risks in functionality and security.
Consequently, it is essential to establish a deprecation mechanism to assist package developers and users in managing packages effectively.
 
To facilitate the establishment of the package-level deprecation mechanism, this paper presents a mixed-method empirical study, including data analysis and surveys. We investigate the current practices of announcing, receiving, and handling package-level deprecation in the Python ecosystem. We also assess the benefits of having deprecation announcements for inactively maintained packages. Furthermore, we investigate the challenges faced by package developers and users and their expectations for future deprecation practices.
Our findings reveal that
75.4\% of inactive package developers have no intention of releasing deprecation declarations for various reasons, while 89.5\% of users express a desire to be notified about the deprecation, highlighting a gap between developers and users;
in many cases, no alternative solutions are available when deprecation occurs, emphasizing the need to explore practical approaches that enable seamless package handover and require less maintenance effort.
Our work aims to enhance the understanding of existing package-level deprecation patterns within the Python OSS realm and facilitate the development of deprecation practices for the Python community in the future.
\end{abstract}

\begin{IEEEkeywords}
Deprecation, open source, empirical study
\end{IEEEkeywords}

\section{Introduction}

Open-source software (OSS) has become a ubiquitous and influential presence in the software industry, resulting in considerable economic ramifications \cite{eghbal2016roads, henry2023measuring, valiev2018ecosystem}. Notably, a recent survey \cite{openuk2021state} uncovered that 89\% of UK companies utilize OSS. 
However, OSS projects are usually developed and maintained by volunteers~\cite{eghbal2016roads, miller2023we}, who may discontinue their involvement due to time constraints, lack of interest, or personnel turnover \cite{coelho2017modern, yu2012empirical}, resulting in inadequate or no maintenance for the software. Adopting poorly maintained OSS projects poses significant risks and potential losses to the software industry. 
Synopsys' recent report on open-source security and risk analysis (OSSRA) \cite{synopsys2023open} revealed that 89\% of the analyzed codebases contained outdated open-source components that were over four years old, and 84\% of them had at least one known open-source vulnerability. 

To mitigate the challenges resulting from outdated code and its associated issues, such as the propagation of vulnerability, researchers have recommended the implementation of a \texttt{deprecation} mechanism~\cite{decan2018impact}. 
The deprecation mechanism allows developers to officially declare the software as obsolete and discourage users from using it~\cite{zhou2016api,cogo2021deprecation}.
Various programming language ecosystems have embraced the concept of deprecation, including languages such as JavaScript and Rust. For instance, in JavaScript, Node Package Manager (NPM) provides a deprecation mechanism that allows developers to mark packages as deprecated~\cite{npmdeprecate}, thereby warning users about their outdated nature and suggesting alternative solutions.

Surprisingly, despite being one of the most popular programming languages~\cite{popular_pl} with over 400k third-party packages, Python lacks a formal or widely recognized package-level deprecation mechanism in its ecosystem, hindering package developers from releasing explicit deprecation announcements.
In this paper, deprecated packages in the Python ecosystem are identified as those that include an explicit deprecation declaration, indicating that they will no longer receive maintenance.
We consider inactive packages as those that have not received any updates for a long time, surpassing their regular maintenance schedule.
It is important to note that not all inactive packages are necessarily abandoned. Some may be considered complete in functionality, thus slowing down their maintenance~\cite{coelho2017modern}.
This intricate nature of inactive packages poses challenges for users in determining the package's status when no deprecation announcement is available, subsequently exposing them to potential risks and issues~\cite{valiev2018ecosystem}.
To mitigate these negative impacts, there is a strong demand to develop a deprecation mechanism in the Python ecosystem that enables developers to deliver explicit and effective information.

While the Python ecosystem does not have an official package-level deprecation mechanism, developers have implemented their own approaches to announce deprecation in recent years. One common practice is to include deprecation information in the README file of the package repository. Furthermore, developers often offer alternative package options as part of the deprecation process. 
However, there is a lack of awareness regarding the existing deprecation patterns throughout the entire Python ecosystem. 
The prevalence of these deprecation patterns is also unknown.
In addition, it is important to investigate whether these existing deprecation patterns meet the requirements of both package developers and users.
To address this gap, we conducted a mixed-method empirical study to examine the early efforts made in deprecating Python packages and gather feedback from package developers and users regarding the current deprecation patterns.
Concretely, we address the following research questions:

\vspace{0.25cm}
\emph{\textbf{RQ1}: How is package-level deprecation currently made, received, and handled?}\vspace{0.10cm}\\
We successfully identified 9,173 deprecated packages that announce their deprecation through various methods, such as GitHub archiving, homepage notifications, issue trackers (e.g., GitHub issues) when users inquire, and warnings during installation.
However, these deprecated packages only account for 1\% of the inactive packages, and for the remaining ones, we are unable to determine their status.
Among the identified deprecated announcements, only 8.7\% have provided alternative solutions.
Even after six months since the deprecation was announced, two-thirds of the users remain unaware of the deprecation.
This is primarily due to infrequent checks on the maintenance status of the dependencies of their packages.
Nevertheless, over two-thirds of the users are willing to take action regarding the dependency when directly or transitively affected by vulnerabilities. This includes removing the deprecation, finding alternative solutions, and creating a new fork.

\vspace{0.25cm}
\emph{\textbf{RQ2}: Can deprecation announcements mitigate the negative impacts of inactive maintenance?}\vspace{0.10cm}\\
We verified that having a deprecation announcement can reduce the unresolved issues on GitHub and the adoption of downstream packages, indicating the alleviation of risks.
Additionally, most users agree that having a deprecation announcement helps reduce the decision efforts when deciding whether to adopt or remove an inactive package.

\vspace{0.25cm}
\emph{\textbf{RQ3}: Why do inactive packages rarely release a deprecation announcement?}\vspace{0.10cm}\\
This is primarily attributed to a lack of time and resources, uncertainty about how to provide an announcement and make future plans.
Moreover, it is worth noting that developers often have limited awareness of the usage of their packages, which further complicates their decision-making process regarding future plans and whether to announce a deprecation.

\vspace{0.25cm}
\emph{\textbf{RQ4}: What are the expectations of package developers and users regarding the future deprecation pattern?}\vspace{0.10cm}\\
While the majority of the package developers expressed their willingness for the package manager to automatically handle deprecation tasks upon their request, announcing the deprecation manually is still preferred by many developers.

As for package users, the majority of them express their need to receive deprecation notifications through warnings during installation or via email, which should also include information about alternative solutions.
They also want additional support, such as 
migration guidelines, and guidance on taking over the projects.
\vspace{0.25cm}

In summary, we made the following contributions:

\begin{itemize}[topsep=5pt]
\item We performed an empirical study to understand the status quo and challenges of package-level deprecation in the Python ecosystem. This study can serve as a cornerstone for establishing a deprecation mechanism for Python.
\item We discussed the practical implications of managing package-level deprecation for the community.
\item We collected and released a large-scale dataset at \url{https://doi.org/10.5281/zenodo.13335360}, consisting of 106,323 packages.
The dataset can be used to facilitate future research.
\end{itemize}

\section{Background}

Inactive packages, also known as dormant~\cite{valiev2018ecosystem}, abandoned~\cite{avelino2019abandonment}, halted~\cite{pashchenko2018vulnerable}, or unmaintained~\cite{coelho2017modern} packages, have not received sufficient maintenance in recent times, posing risks to users who integrate the packages into their projects.
The lack of maintenance can lead to package incompatibility with new modules or environments over time, thereby impacting the functionality.
When users seek assistance by submitting an issue, it is likely that they will receive no support from the package developers.
Even if the functionality remains unchanged, packages can still be affected by vulnerabilities from their dependency silently. 
For example, in 2021, the discovery of a severe vulnerability in Log4J caused widespread panic in the entire Internet industry as many software directly or transitively adopted this library.
Although the vulnerability was patched rapidly, the propagation of these patches to downstream packages can be hindered by inactive packages in the software supply chain. Consequently, the vulnerability remains a persistent risk for years~\cite{IBM2023Log4j,wang2023plumber}.

In general, developers utilize a deprecation mechanism to discourage consumers from adopting obsolete functionalities.
A typical deprecation mechanism follows three steps. The first step is declaration, where the deprecation status is explicitly stated. The second step is to deliver the deprecation message to the consumers. Lastly, index update prevents consumers from searching for deprecated packages or releases.

Depending on the objects or functionalities to be deprecated, different levels of deprecation mechanisms exist.
Traditionally, developers can leave warnings with an \textit{API-level} deprecation mechanism to discourage consumers from adopting an abandoned object or functionality~\cite{java_depre_doc, PEP702}.
With the evolution of software, numerous historical releases remain in the ecosystem and are available for consumers to choose from. If specific versions of package releases are found to be buggy or even vulnerable, developers may consider using a \textit{release-level} deprecation mechanism to deprecate those specific versions~\cite{PEP592}.
However, developers who adopt these practices do not necessarily intend to give up the maintenance of entire packages.
\begin{figure}[h]
    \centering
    \includegraphics[width=0.5\textwidth]{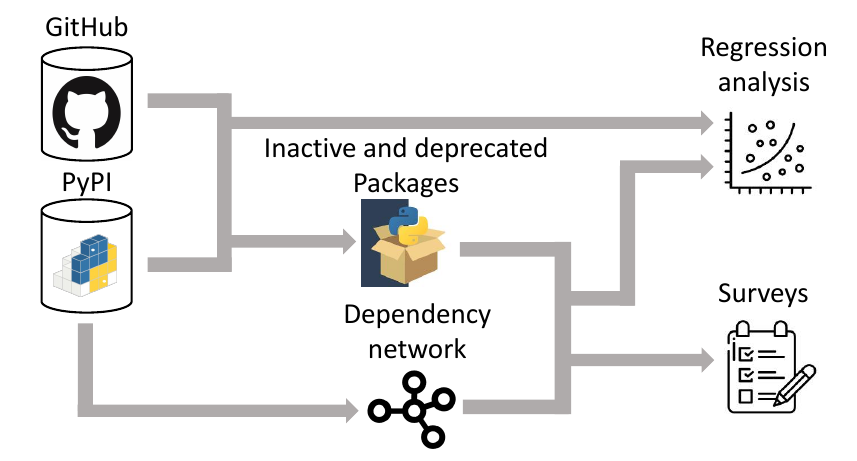}
    \caption{The methodology of our study}
    \label{fig:flow chart}
\end{figure}
In cases where developers decide to give up maintaining packages and discourage consumers from further using the package, a \textit{package-level} deprecation mechanism can be employed to deprecate the unmaintained packages and deliver the deprecation declaration.
Various programming languages already have existing package-level deprecation mechanisms.
For instance, in npm for Javascript, developers can implement a package-level deprecation mechanism to leave a message for new consumers attempting to install a deprecated package.


However, despite discussions regarding a package-level deprecation mechanism for the Python ecosystem for years~\cite{naveen2021deprecate,Gedam2022adding}, no existing package-level deprecation mechanism currently exists within the Python ecosystem.
The absence of the deprecation mechanism leaves package users unaware of the risks associated with adopting the package, thereby increasing the decision effort required.
One motivating example is \texttt{Python-Kafka}, which has no release in the last 1.5 years.
Users are concerned about the compatibility risks of using the package and have already encountered bugs. As a result, they have inquired about whether the package has been deprecated via submitting a GitHub issue. However, the reply remains ambiguous, leaving users in a state of uncertainty~\cite{rj32022kafka}.
To facilitate the establishment of a deprecation mechanism in the Python ecosystem, we conducted a mixed-method empirical study that included data analytics and two user studies.


\section{Methodology}
\label{sec:Methods}
Figure~\ref{fig:flow chart} illustrates the overview of our methodology for addressing the research questions.
Initially, we collected data from PyPI and GitHub, allowing us to identify both inactive packages and deprecated packages. 
Next, we extracted the dependency relationships among packages from the collected data.
Subsequently, we analyzed the benefits of deprecation announcements via regression analyses.
To delve deeper into the practices, challenges, and expectations regarding package-level deprecation in the Python ecosystem, we designed two surveys from inactive package developers and deprecated package users and collected their feedback.

\subsection{Data Collection}
\label{sec:data_collection}
\subsubsection{PyPI}
We initially obtained the metadata for all 415,151 available packages using the official PyPI API as of January 09, 2023, where the release history and URLs are used in the follow-up analyses (Section~\ref{sec:dependency_network}, \ref{sec:identifing}).
Furthermore, we downloaded every release for each package to extract the dependency information~\cite{valiev2018ecosystem}.



\subsubsection{GitHub Data}
To gather more data about PyPI packages, we collected additional information from their corresponding repositories on GitHub.
Following a similar procedure to prior work~\cite{valiev2018ecosystem}, we initially mined the relationship between PyPI packages and their repositories.
Specifically, we first examined three metadata fields: \textit{homepage}, \textit{repository}, and \textit{project\_urls}, to identify URLs that match popular code hosting platforms (e.g., github.com, gitlab.com).
In cases where no URL was found, we subsequently searched all other metadata fields for URLs containing the package name.
If a URL was not still found, we downloaded the latest package release and searched for URLs in all package files that contained the package name.
Overall, out of the 415,151 packages on PyPI, we successfully identified 300,175 URLs (72.3\%), with 287,624 (69.3\%) of them originating from GitHub, which is consistent with the earlier report~\cite{valiev2018ecosystem}.
Since most of the identified URLs are from GitHub, we focused our data mining efforts on packages associated with GitHub repositories for simplicity.

After establishing the relationship between PyPI packages and their corresponding GitHub repositories, we utilized the GitHub REST API~\cite{github_rest_api} to collect various data, such as README, open issues, and commit histories. 

\subsection{Dependency Network}
\label{sec:dependency_network}
To analyze the benefits of deprecated packages, we constructed a dependency network for the whole PyPI ecosystem.
Our analysis focuses on packages explicitly declared as dependencies in other PyPI packages.  
This excludes dependencies outside the PyPI ecosystem, such as system dependencies and GitHub projects unrelated to PyPI. 
In the PyPI package context, we refer to packages used by the current package as ``upstream packages" or ``upstream dependencies". Conversely, ``downstream packages" or ``downstream dependencies" are packages that depend on the current package. 

In order to ensure comprehensive and up-to-date dependency information, we extracted this data from distribution files of each version, following the approach of prior work~\cite{valiev2018ecosystem}.
PyPI supports two types of package distributions: built distribution and source distribution.
Initially, we prioritized the built distribution as it contains machine-readable metadata for dependencies.
If built distributions were unavailable, we performed a mimic installation using the source distribution in a sandbox environment.
Throughout this process, we logged the dependencies required for installation. 
In total, we obtained dependency information from 4,034,194 releases.

\subsection{Identifying Inactive Packages}
\label{sec:identifing}
\def\VarName{2.3cm}
\def\ChaName{2.7cm}
\def\OutLen{6cm}
\def\DescLen{5cm}
\begin{table*}[h]
  \centering
  \caption{The definitions of the variables in our models}
  \label{tab:definitions}
  \begin{tabular}{llll}
    \toprule
    \multicolumn{2}{l}{\textbf{\textit{Outcome Variables}}} & \multicolumn{2}{l}{\textbf{\textit{Package Characteristics}}}\\
    \multirow{12}{\VarName}{\parbox{\VarName}{\vspace{.5\baselineskip}Gain of downstream dependencies\vspace{.5\baselineskip}}} & \multirow{8}{\OutLen}{
     \parbox{\OutLen}{We define the gain of downstream dependencies in PyPI as follows: if the current package is solely depended on by the latest version of a downstream package, we consider it an increase in downstream dependencies. Conversely, if there are versions, other than the latest one, that depend on the current package, we consider it a decrease in downstream dependencies. The gain of downstream dependencies is calculated as the difference between the increase and decrease in downstream dependencies. }} & & \\
           & &\parbox{\ChaName}
     {\vspace{.5\baselineskip}Is deprecated \vspace{.5\baselineskip}}  & 
       \parbox{\DescLen}{A boolean variable indicating whether the package is deprecated or not.\vspace{.5\baselineskip}}\\
       & & \parbox{\ChaName}{\vspace{.5\baselineskip}Number of downstream dependencies ever\vspace{.5\baselineskip}}  & 
       \parbox{\DescLen}{\vspace{.5\baselineskip}The number of downstream packages that have either adopted or previously adopted the current package.\vspace{.5\baselineskip}} \\
       & & {\vspace{.5\baselineskip}Project age \vspace{.5\baselineskip}}  & 
       \parbox{\DescLen}{\vspace{.5\baselineskip}The duration in days from the first commit to the last commit.\vspace{.5\baselineskip}} \\
       &  & \parbox{\ChaName}
     {\vspace{.5\baselineskip}Project stars \vspace{.5\baselineskip}}  & 
       \parbox{\DescLen}{\vspace{.5\baselineskip}The number of stars received by the corresponding GitHub repository.\vspace{.5\baselineskip}}\\
       \multirow{7}{\VarName}{\parbox{\VarName}{\vspace{.5\baselineskip}Gain of unresolved issues\vspace{.5\baselineskip}}} & \multirow{5}{\OutLen}{
     \parbox{\OutLen}{This variable refers to the number of GitHub issues that are in an open state and have not received any reply from any official member of the repository. Specifically, we consider issues that were created after the last commit to the repository. }} & \parbox{\ChaName}
     {Number of contributors }   & 
       \parbox{\DescLen}{The number of contributors involved in the corresponding GitHub repository.}\\
       & & \parbox{\ChaName}
     {\vspace{.5\baselineskip}Number of releases \vspace{.5\baselineskip}}  & 
       \parbox{\DescLen}{\vspace{.5\baselineskip}The number of releases uploaded to PyPI.\vspace{.5\baselineskip}} \\
       & & \parbox{\ChaName}
     {\vspace{.5\baselineskip}Duration since last commit \vspace{.5\baselineskip}}  & 
       \parbox{\DescLen}{\vspace{.5\baselineskip}The number of days from the last commit to the present (as of 2023.3.27).\vspace{.5\baselineskip}}  \\

  \bottomrule

\end{tabular}
\end{table*}
To facilitate understanding the maintenance status of the packages in the PyPI ecosystem, and identifying the deprecated packages,
we first identified inactive packages from the entire set of 415,151 packages on PyPI, following the approaches used in prior works.
There are two main types of methods used to identify inactive packages.
One type of method considers a specific number of commits within a defined duration, such as no commit in the last year, as criteria for determining inactive packages~\cite{khondhu2013all,coelho2017modern, kula2018developers,valiev2018ecosystem, ait2022empirical}.
The other type of method identifies packages as inactive if they exceed the expected time to release a new version based on their release history~\cite{pashchenko2018vulnerable,pashchenko2019decision, pashchenko2020vuln4real}.
However, both of these methods can result in many false positives.
For mature packages, it may take a long time to release a new version.
While for those new packages, their releasing schedule may be irregular.
To minimize false positives, we only considered packages as inactive if \textit{both} methods identified them as such.

Specifically, for the first type of method, we adopted the widely used criteria of identifying packages without any commits in the last year.
For the second type of method, we employed an Exponential Smoothing model~\cite{brown1959statistical} to calculate the next expected release date, following the approach outlined in prior work~\cite{pashchenko2018vulnerable}.
The model is presented in Equation~\ref{eq:next_release_date},
where $\alpha$ represents the smoothing parameter that takes a value of 0.6, as consistent with prior work~\cite{pashchenko2018vulnerable}. $n$ denotes that there are $n$ intervals in the release history (i.e., $n+1$ releases), while $Release\ time_i$ denotes the time required to release a new version after the \textit{i}-th version.
This model assigns greater weight to more recent release intervals compared to older intervals.
By combining these two methods, we identified a total of 103,733 unique inactive packages within the ecosystem.
\begin{equation}
\label{eq:next_release_date}
\small 
\begin{aligned}
Next\ release\ interval &= \alpha \sum^{n-1}_{i=0}\{(1-\alpha)^i\ast Release\ time_{n-i-1}\} \\ 
Expected\ release\ date &= Next\ release\ interval \\
&+ Last\ release\ date
\end{aligned}
\end{equation}
\subsection{Identifying Deprecated Packages}
\label{sec:identifying_depre}
To identify deprecated packages, we first collected packages labeled as ``archived" on GitHub. The ``archive'' feature allows GitHub developers mark a package as obsolete and discourage its use~\cite{GitHub2021Archiving}, which aligns with the definition of deprecation in this paper.
We used a web crawler to check the corresponding GitHub repositories of PyPI packages and determine if they were archived on GitHub.
Through this process, we successfully identified 8,932 packages that had been archived.

In addition, we labeled inactive packages that contain a deprecation announcement as deprecated, resulting in 1,596 inactive packages.
To achieve so, we initially conducted keyword searches for a set of commonly used phrases (e.g., deprecated, unmaintained, not maintained anymore, etc.) in the README files, issues created after the last commit, and the setup program in the PyPI distributions, following prior work~\cite{coelho2018identifying}.
If any of these phrases were found, two of our authors independently performed manual validation to determine whether the identified keywords indicated package deprecation.
The inspection results achieved a Cohen's kappa score of 0.846, indicating an almost perfect inter-rater agreement~\cite{mchugh2012interrater}. 
We also collected the rationales and alternative solutions if available.
In total, we have identified 9,173 deprecated packages.

\subsection{Estimating Effect of Deprecation Announcement}
\label{sec:effect}
We performed a regression analysis to examine whether a deprecation declaration can alleviate the negative impacts of inactive packages.
The unit of analysis is a package that is either deprecated or just inactive.
All variables we adopted in the regression analysis and their definitions are listed in Table~\ref{tab:definitions}. 
For every package, we leveraged data that was collected in Section~\ref{sec:data_collection} and Section~\ref{sec:dependency_network} to capture the characteristics of the packages.
Given that unresolved GitHub issues and downstream dependencies are commonly considered as negative impacts of inactive packages~\cite{coelho2017modern,jiang2017understanding}, we also include these two variables as the dependent variables in the regression analysis.


\subsubsection{Model Specification} We utilized Linear Mixed Effect Regression (lmer) models to estimate the impact of having a deprecation announcement, which is demonstrated in Equation~\ref{eq:deprecation}.
Specifically, we want to estimate whether the deprecation will lead to the change of unresolved issue amount (Model I). Besides, we try to understand whether it could also affect the number of downstream dependencies (Model II).
\begin{equation}
\label{eq:deprecation}
\begin{aligned}
Y = \beta_0d  + \beta_1r+ \delta  
\end{aligned}
\end{equation}
Here, $d$ is a boolean variable indicating whether the package is deprecated or not,
and $r$ denotes other repository characteristics outlined in Table~\ref{tab:definitions}.
The coefficients of the corresponding variables are denoted as $\beta_i$,
and the ``owner-cohort" random effect is represented by $\delta$. 
$Y$ represents the outcome variables. 



\subsubsection{Model Estimation}
\label{sec:estimation}
Before estimating the regression, we carefully selected subsets of inactive packages and deprecated packages to minimize bias.
We utilized a sample size calculator~\cite{Sample_size} that is widely adopted by many prior SE works~\cite{asyrofi2021biasfinder,wang2021restoring} to determine the sample size with statistical significance. For Model I, we sampled 369 deprecated packages.
As for Model II, we sampled 226 deprecated packages from deprecated packages that announce their deprecation through methods except for GitHub archiving.
This is because once a package is archived on GitHub, all the open issues are closed, and no more new issues can be submitted to the repository.
Next, we traversed the dependency network for each deprecated package and identified all its brother packages (i.e., those that share at least one common dependency) among all inactive and deprecated packages. This step was taken to manage risks associated with the upstream dependencies. For example, these packages may become inactive as a result of the same incident occurring in the upstream dependency.
Rather than including all its brothers in the subset, we only selected 5 packages with the closest number of stars in the corresponding GitHub repository.

Moreover, we took standard precautions while estimating the models in the following manner.
We used log-transforming variables to mitigate heteroscedasticity with skewed distributions \cite{gelman2006data}, except for the gain of downstream dependencies, which includes negative values.
To assess multicollinearity, we utilized the \textit{Variance Influence Factor} (VIF) from the \textit{car} package in R~\cite{fox2019companion}.
The results are presented in Table~\ref{tab:results}.



\subsection{Survey for Developer and User}
To gain an in-depth understanding of the practices and demands of both package developers and users in the ecosystem when encountering deprecation, we conducted two separate surveys for package developers and users.
Similar to prior work, we used Microsoft Forms~\cite{Mircrosoft2021forms} to design two survey questionnaires~\cite{he2022empirical}.
We then separately sent our invitation emails to package developers and users, providing them with the purpose of the survey and a link to the respective questionnaire.
After that, two of our authors analyzed the results following the standard procedure.
\begin{table}[h!]  
\begin{center}  
\def\CatLength{1.75cm}
\def\QuesLength{5cm}
\caption{Survey for inactive package developers}  
\label{tab:questionnaire1}  
\begin{tabular}{c|c|l}  
  
\toprule 
  
 & \textbf{ID} & \textbf{Question}\\  
  
\midrule 
   
 & Q1 & \parbox{\QuesLength}{\vspace{.25\baselineskip} What's the maintenance status of the package?\vspace{.25\baselineskip}}\\  
  \parbox{\CatLength}{\vspace{.25\baselineskip} \centering Maintenance Status\vspace{.25\baselineskip}}& Q2 &\parbox{\QuesLength}{\vspace{.25\baselineskip} What are the reasons for not maintaining the package?\vspace{.25\baselineskip}}\\  
  & Q3 & \parbox{\QuesLength}{\vspace{.25\baselineskip}Do you know any alternative solution or package?\vspace{.25\baselineskip}}\\  
\midrule
  \multirow{7}{1.5cm}{\centering Practices and Challenges}  & Q4 & \parbox{\QuesLength}{\vspace{.25\baselineskip} Do you intend to announce the deprecation in the future, with the option to withdraw it if you decide to restart the package?\vspace{.25\baselineskip}}\\  
 & Q5 & \parbox{\QuesLength}{\vspace{.25\baselineskip} What's the difficulty that prevents you from providing an explicit announcement?\vspace{.25\baselineskip}}\\  
 & Q6 & \parbox{\QuesLength}{\vspace{.25\baselineskip} If you want to announce the deprecation, how would you like to make the announcement?\vspace{.25\baselineskip}}\\  
  
\bottomrule 
  
\end{tabular}  
\end{center}  
\end{table}
In the following sections, we provide details about participant selection, study design, and the analysis of the obtained results.
\subsubsection{Participant Selection}
From the collected results of Section~\ref{sec:identifying_depre}, we found that only a small proportion of the inactive packages have released a depreciation announcement.
For the remaining inactive packages, we are curious about their maintenance status, as well as the current practices and preferences of their developers when it comes to deprecation.
Therefore, we sampled 2,000 packages from inactive packages that did not release deprecation announcements and designed a questionnaire for their developers.
We obtained the email addresses of their developers from the package metadata and sent them an invitation email.
Eventually, we received 118 valid responses.

Once deprecation occurs, we aim to understand how users of deprecated packages perceive and handle deprecation.
To achieve this, we leveraged the dependency network to identify downstream packages that directly depend on deprecated packages.  
Similarly, we sampled 2,000 of those downstream packages that have adopted deprecated packages and designed a questionnaire for their developers.
To gather more valuable insights from users, we ensure that the deprecated package they adopted has released its deprecation announcement for at least six months.
So that the users are more likely to have a deeper understanding and more thoughts about the deprecation.
Eventually, we received 106 valid responses.

\subsubsection{Survey Design} 

\noindent\textbf{Survey for Inactive Package Developers.}
The survey for inactive package developers consists of seven questions, six of which are closed-ended and listed in Table~\ref{tab:questionnaire1}.
The remaining question, an optional open-ended one, is not included in Table~\ref{tab:questionnaire1}.
This optional question aims to gather advice on the survey and any additional thoughts that participants want to share.
The options for the six questions are provided based on our domain experience.
Except for Q3, each question includes an ``Others" option, allowing participants to provide a free-text response if their desired answer is not among the provided options.
\begin{table}[h!]  
\begin{center}  
\def\CatLength{1.75cm}
\def\QuesLength{5cm}
\caption{Survey for inactive package users}  
\label{tab:questionnaire2}  
\begin{tabular}{c|c|l}  
  
\toprule 
  
 & \textbf{ID} & \textbf{Question}\\  
  
\midrule 
   
 Awareness & Q1 & \parbox{\QuesLength}{\vspace{.25\baselineskip}  Do you know that there is a deprecated package in your codebase?\vspace{.25\baselineskip}}\\  
\midrule
  \multirow{3}{1.5cm}{\centering Practices for Checking Deprecation}& Q2 &\parbox{\QuesLength}{\vspace{.25\baselineskip} How often do you check for deprecated dependencies in your projects?\vspace{.25\baselineskip}}\\  
  & Q3 & \parbox{\QuesLength}{\vspace{.25\baselineskip}How do you check whether one package is deprecated or not?\vspace{.25\baselineskip}}\\  
\midrule
  \multirow{7}{1.5cm}{\centering Expectations on Deprecation Pattern}  & Q4 & \parbox{\QuesLength}{\vspace{.25\baselineskip} Do you want to be notified when there are deprecated dependencies in your codebase by the package manager (i.e., PyPI) or package owner?\vspace{.25\baselineskip}}\\  
 & Q5 & \parbox{\QuesLength}{\vspace{.25\baselineskip} How would you expect to be notified about the deprecation of a package?\vspace{.25\baselineskip}}\\  
 & Q6 & \parbox{\QuesLength}{\vspace{.25\baselineskip} What content do you expect to be included in the deprecation announcement?\vspace{.25\baselineskip}}\\  
\midrule
 Willingness& Q7 & \parbox{\QuesLength}{\vspace{.25\baselineskip} Is it more difficult to make the decision without an explicit deprecation announcement regarding whether to adopt, remove, or take other actions on an inactively maintained package that has not been updated for a long time?\vspace{.25\baselineskip}}\\  
\midrule
 \multirow{6}{1.5cm}{\centering Practices of Taking actions} & Q8 & \parbox{\QuesLength}{\vspace{.25\baselineskip} In what situation will you take actions on deprecated packages?\vspace{.25\baselineskip}}\\  
 & Q9 & \parbox{\QuesLength}{\vspace{.25\baselineskip} What actions will you take on the deprecated dependency?\vspace{.25\baselineskip}}\\  
 & Q10 & \parbox{\QuesLength}{\vspace{.25\baselineskip} Are there any challenges when taking action on the deprecated packages?\vspace{.25\baselineskip}}\\  
\midrule
\multirow{1}{1.5cm}[1.3ex]{\centering Further Expectations} & Q11 & \parbox{\QuesLength}{\vspace{.25\baselineskip} What kind of support do you expect from developers and package managers?
\vspace{.25\baselineskip}}\\
\bottomrule 
  
\end{tabular}  
\end{center}  
\end{table} 
In addition, Q2, Q5, and Q6 allow participants to choose more than one option.

The first three questions mainly focus on the maintenance status of inactive packages.
In particular, the first question asks whether the package will continue to receive maintenance in the future.
If the participant confirms that the package is no longer maintained, 
we then inquire about the reasons behind this decision and alternative solutions in Q2 and Q3.
Considering that only a few inactive packages choose to release a deprecation announcement, and even when such an announcement exists, the approaches to make it vary.
Therefore, we are interested in understanding the current practices followed when handling the inactively maintained packages.
Consequently, we asked about their willingness to release a deprecation announcement and the potential difficulties they encountered in Q4 and Q5.
Additionally, we investigate their preferred methods for announcing deprecation in Q6.
This feedback will provide valuable insights for developing a package-level deprecation mechanism in the future.

\noindent\textbf{Survey for Deprecated Package Users.} 
The survey for deprecated package users consists of twelve questions, eleven of which are closed-ended and presented in Table~\ref{tab:questionnaire2}.
We also included an optional question for advice and additional thoughts, which is consistent with the survey for deprecated package developers.
The options for the eleven questions were provided based on our domain experience.
For every question, we also included an ``Others" option.
Apart from Q1, Q2, Q4, and Q7, participants can select multiple options for the remaining questions.
In addition, we introduced some preliminary concepts before presenting the questions to the participants, including our definition of deprecated packages, deprecation announcements, risks of using deprecated packages, and common actions taken on the deprecated packages. This context was intended to enhance participants' understanding of the subsequent questions.

Considering the diverse approaches to announcing the deprecation announcement, we wonder whether these announcements can efficiently reach its users.
Therefore, we investigate users' awareness of the deprecation in Q1.
Then, we investigate what practices users adopt to get themselves notified about the maintenance status of their dependencies in Q2 and Q3.
In Q4, Q5, and Q6, we also collect users' expectations on methods of being notified about the deprecation if they want.
We believe that the feedback can be complementary insights that originate from the point of view of package users.
Since taking actions on the deprecated dependencies consumes time and effort, users may not always choose to take actions. 
We investigate whether a deprecation announcement helps users better make decisions on whether to take actions in Q7.
Then, we investigate under what situations users would take actions in Q8.
In Q9 and Q10, we specifically investigate the practices and potential challenges they would encounter when taking action on the deprecated packages.
Additionally, we investigate their potential additional needs in Q11.


\subsubsection{Organizing Survey Results}

To analyze the survey results, we followed the survey practices presented in recent studies~\cite{alsuhaibani2021naming, he2022empirical}.
Specifically, for closed-ended questions, we summarized the responses by plotting the distribution or counting the number of participants who selected each option when multiple options were allowed. 
Regarding the free-text responses collected through the ``Other" option, we employed widely adopted open coding methods for summarizing the results~\cite{cruzes2011recommended}, in line with recent studies~\cite{alsuhaibani2021naming,he2022empirical}.
Specifically, two of our authors independently went through each response and assigned preliminary labels.
Then they reviewed the responses again to refine the labels.
Finally, the two authors discussed their refined labels and compiled a final list of the labels.
With this final list, we proceeded to relabel each response accordingly. 
If both authors agreed that certain responses were actually pre-existing options that we have provided in the questionnaire, we labeled those responses accordingly.
We also excluded any responses that were not relevant to the questions.

\begin{figure}[h]
    \centering
    \includegraphics[width=0.5\textwidth]{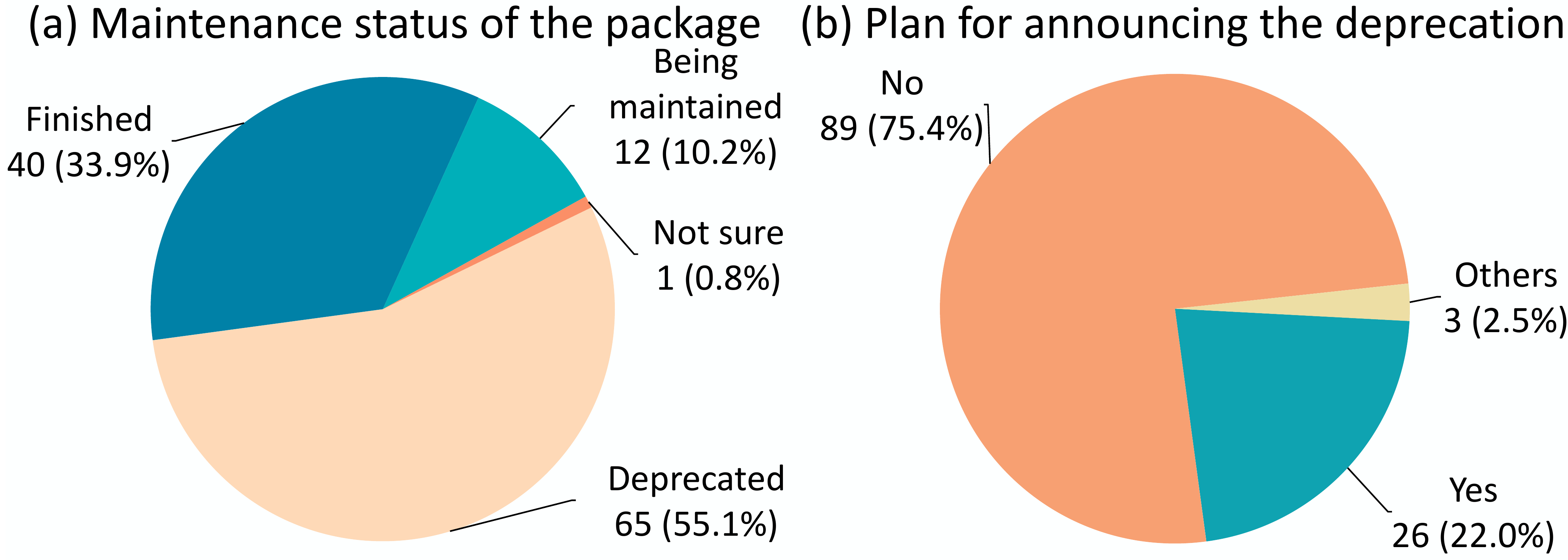}
    \caption{Results of Q1 and Q4 in the Survey for inactive package developers}
    \label{fig:q1s1_q4s1}
\end{figure}
\subsection{Ethics}
To minimize unnecessary spamming, when designing the surveys, we specifically targeted people who might be interested in participating, including inactive package developers and deprecated package users. We only sent one invitation email to the potential participants rather than multiple emails to everyone. In the email, we explained that the purpose of the surveys is to understand their needs and expectations, as well as to promote the establishment of a package-level deprecation mechanism for the community. From the responses, we received some expressions of gratitude from participants for notifying them about the deprecated packages and for the study. They also asked to be informed about the study’s progress and findings once it is completed and prepared for public release.
\section{Results}

\subsection{RQ1: How is package-level deprecation currently made, received, and handled?}
\subsubsection{How is package-level deprecation made by developers}
\label{sec:how_deprecations_been_made}
We answer this question by investigating the incidence of deprecation, the approaches employed, and the content included in the announcements.

We identified a total of 9,173 distinct deprecated packages, which consist of packages archived on GitHub and inactive packages with deprecation announcements. Notably, the deprecated packages make up only 1\% of the total inactive packages.
Nevertheless, as Figure~\ref{fig:q1s1_q4s1}(a) depicts over half of the inactive package developers acknowledge that their packages are indeed deprecated. Figure~\ref{fig:q1s1_q4s1}(b) reveals that most of them do not plan to make a deprecation announcement, indicating a low willingness to announce deprecation.

For the methods employed to announce deprecation, it was found that 98.9\% of the deprecation announcements can be accessed on the homepage, typically through the presence of an archived banner or a notification on the README file.
In addition, 1.9\% of the announcements can only be accessed via checking the issues.
0.1\% of the deprecation announcement can be found within setup programs.

Regarding the contents included in the deprecation announcements, it was found that only 2.9\% of them contain the rationales for the deprecation.
The most common rationales identified are the lack of time and resources,
and the existence of better alternatives, which aligns with the findings from Q2 in Table~\ref{tab:questionnaire1} and prior work~\cite{coelho2017modern}.
\begin{figure}[h]
    \centering
    \includegraphics[width=0.5\textwidth]{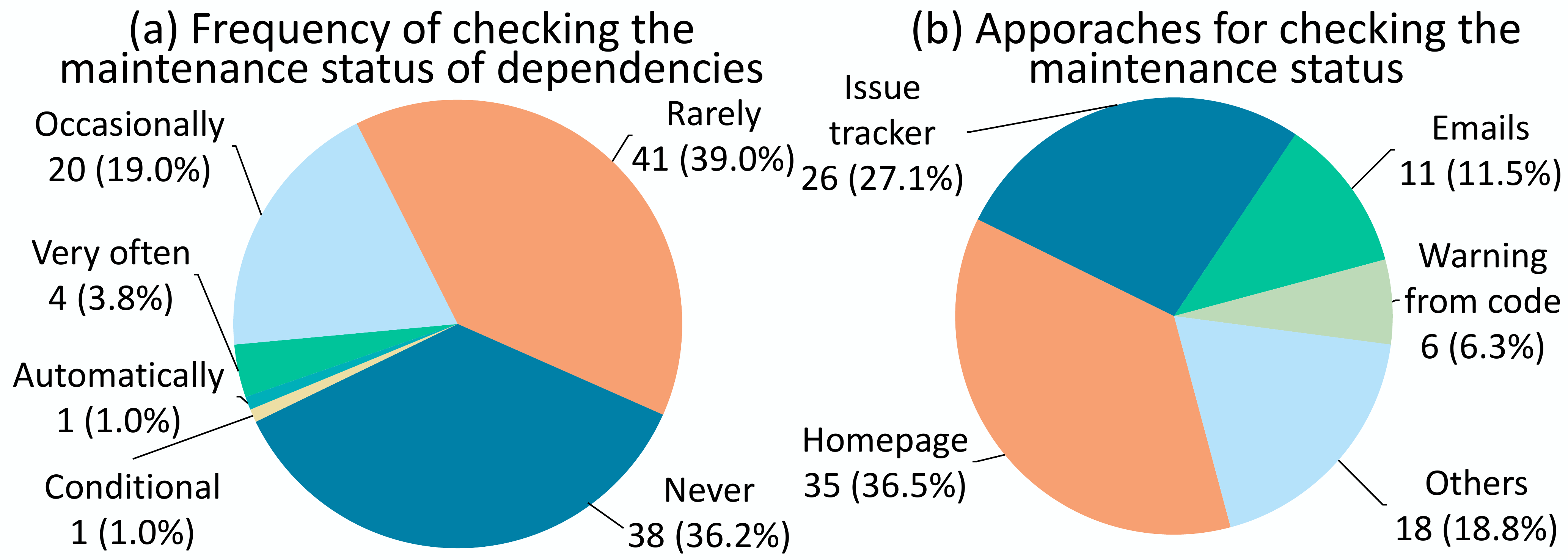}
    \caption{Results of Q2 and Q3 in the Survey for deprecated package users}
    \label{fig:q2s2_q3s2}
\end{figure}
According to the results of Q3 in Table~\ref{tab:questionnaire1}, 39.1\% of the respondents are aware of alternative solutions for their deprecated packages.
However, among the identified deprecated packages, only 8.7\% of them provide an alternative solution, which is even lower.

\begin{tcolorbox}[boxsep=1pt,left=2pt,right=2pt,top=3pt,bottom=2pt,width=\linewidth,colback=white!90!black,boxrule=0pt, colbacktitle=white!,toptitle=2pt,bottomtitle=1pt,opacitybacktitle=0]
\textbf{Finding 1.} \textit{Despite the high possibility that an inactive package is indeed deprecated, only a small number of developers are willing to announce the deprecation, leading to a scarcity of deprecation announcements. Most of the identified deprecation announcements can be accessed on the respective package's homepage. Moreover, only a few participants in our survey are aware of alternative solutions, and the announcements that actually contain alternative solutions are even fewer.}
\end{tcolorbox}
\subsubsection{How is deprecation received by package users}
\label{sec:receiving_depre}
To answer this question, we mainly focus on the responses from the survey for deprecated package users. We specifically investigated their awareness of deprecation and the practices they adopted to perceive deprecation.
Based on the results obtained from Q1 in Table~\ref{tab:questionnaire2}, we found that even after the deprecation has been announced for over six months, two-thirds of the deprecated package users remain unaware of the deprecation.
We further investigated their frequency and methods of checking the maintenance status of the dependencies of their packages in Q2 and Q3.
As Figure~\ref{fig:q2s2_q3s2} depicts, over two-thirds of the users rarely or never check the maintenance status.
When users do decide to check, the most commonly adopted approach is to review the homepage or the issue tracker.
Moreover, from the responses of Q4 in the same survey, we found that most (89.5\%) users are willing to be notified about the deprecation.

\begin{tcolorbox}[boxsep=1pt,left=2pt,right=2pt,top=3pt,bottom=2pt,width=\linewidth,colback=white!90!black,boxrule=0pt, colbacktitle=white!,toptitle=2pt,bottomtitle=1pt,opacitybacktitle=0]
\textbf{Finding 2.} \textit{Two-thirds of deprecated package users are unaware of their dependencies' deprecation, likely due to infrequent checks on the maintenance status. However, it is worth noting that most users are willing to be notified about the deprecation.}
\end{tcolorbox}

\subsubsection{How is deprecations handled}
To answer this question, we also focus on the responses from the surveys for deprecated package users. We specifically investigated the circumstances in which package users take action on the deprecated packages and the specific actions they would take.
\begin{figure}[h]
    \centering
    \includegraphics[width=0.5\textwidth]{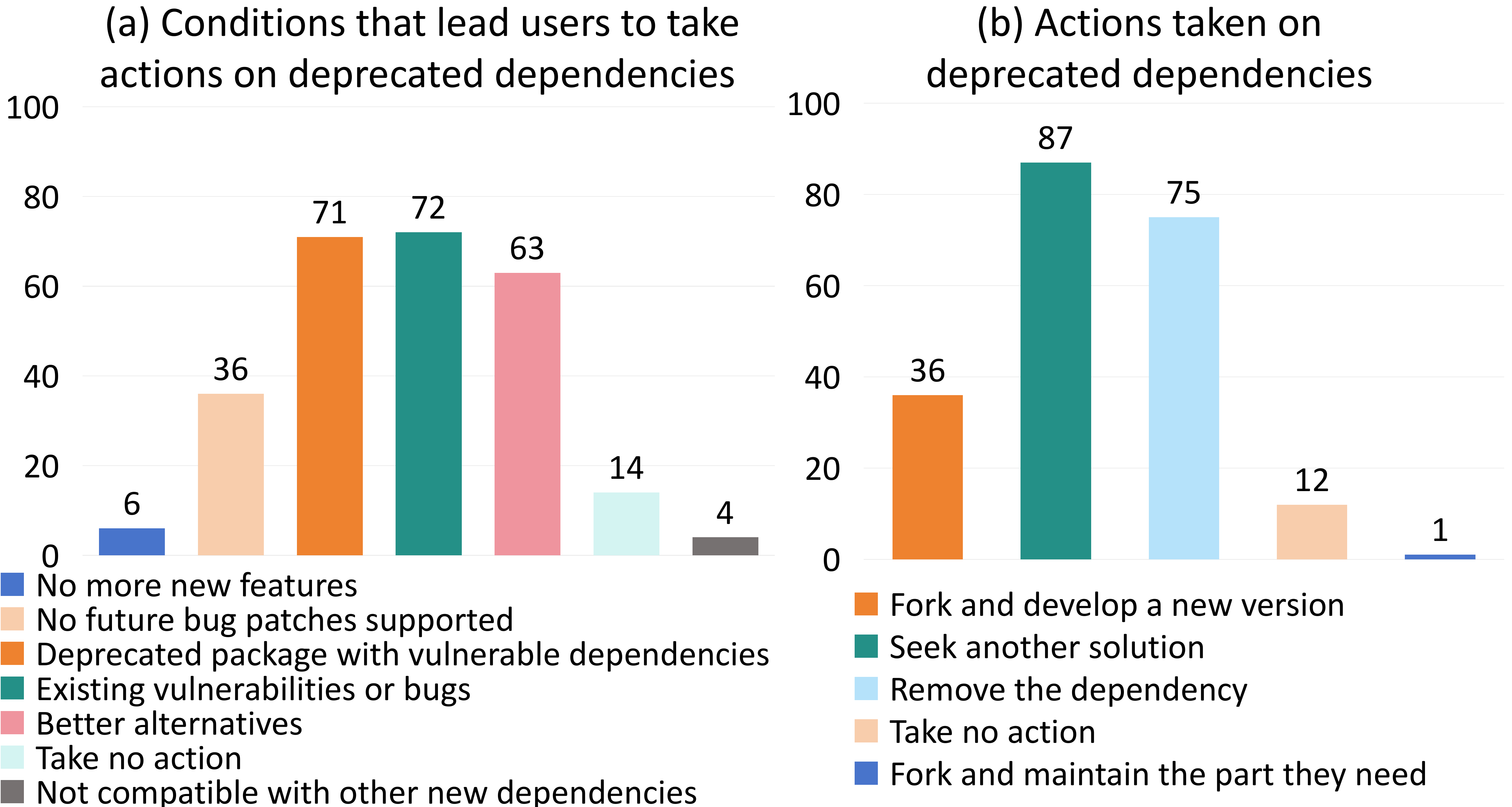}
    \caption{Results of Q8 and Q9 in the Survey for deprecated package users}
    \label{fig:q8s2_q9s2}
\end{figure}
As depicted in Figure~\ref{fig:q8s2_q9s2}(a), vulnerability-related cases are the most commonly mentioned circumstances triggering users to take action.
Besides, 59.4\% of the users expressed that they would take action if better alternatives were available.
Furthermore, 34.0\% of users mentioned that they would take action when future bug patches are no longer provided.
Moving on to the specific actions users would take, as explored in Q9,
Figure~\ref{fig:q8s2_q9s2}(b) reveals that 82.1\% of the users will seek alternative solutions, while 70.8\% of them will remove the dependencies. Moreover, 34.0\% mentioned they will fork the old project and develop a new version.
However, when considering the challenges associated with taking action, the responses from the Q10 indicate that a majority of users cited a lack of time and resources as significant obstacles.

\begin{tcolorbox}[boxsep=1pt,left=2pt,right=2pt,top=3pt,bottom=2pt,width=\linewidth,colback=white!90!black,boxrule=0pt, colbacktitle=white!,toptitle=2pt,bottomtitle=1pt,opacitybacktitle=0]
\textbf{Finding 3.} \textit{Two-thirds of users are willing to take action on the deprecated dependencies in cases involving vulnerability-related issues or the availability of better alternatives. The most common actions taken include removing the dependency and seeking another solution. In addition, one-third of users would like to fork the project and develop a new version.}

\end{tcolorbox}

\subsection{RQ2: Can deprecation announcements mitigate the negative impacts of inactive maintenance?}
\label{sec:estimating_impact}
To answer this question, we verified the benefits of having a deprecation announcement from both quantitative and qualitative perspectives. Quantitatively, we leveraged the regression models estimated in Section~\ref{sec:effect} to verify the impact of such deprecation announcements.
From a qualitative perspective, we analyzed the results of Q7 in Table~\ref{tab:questionnaire2}.

%
Table~\ref{tab:results} presents the estimated coefficients and standard deviations of predictors in two models.
Starting with Model I, the results indicate that a deprecation announcement has a statistically significant impact on reducing the gain of downstream dependencies.
On average, having a deprecation declaration corresponds to approximately 1.15 downstream package decreases.
It is important to note that our observations may not include numerous downstream projects not listed on PyPI, such as those hosted on GitHub.
Moving to Model II, we analyze the impact of preventing the gain of unresolved issues.
Similarly, a deprecation announcement also demonstrates a statistically significant impact on reducing the gain of unresolved issues. 
On average, it corresponds to a decrease of approximately a 17\% in unresolved issues.
From a qualitative standpoint, the results of Q7 in Table~\ref{tab:questionnaire2} reveal that the majority (81.0\%) of package users agree that the absence of a deprecation announcement can increase difficulties in determining how to handle the inactive packages.
\begin{figure}[h]
    \centering
    \includegraphics[width=0.5\textwidth]{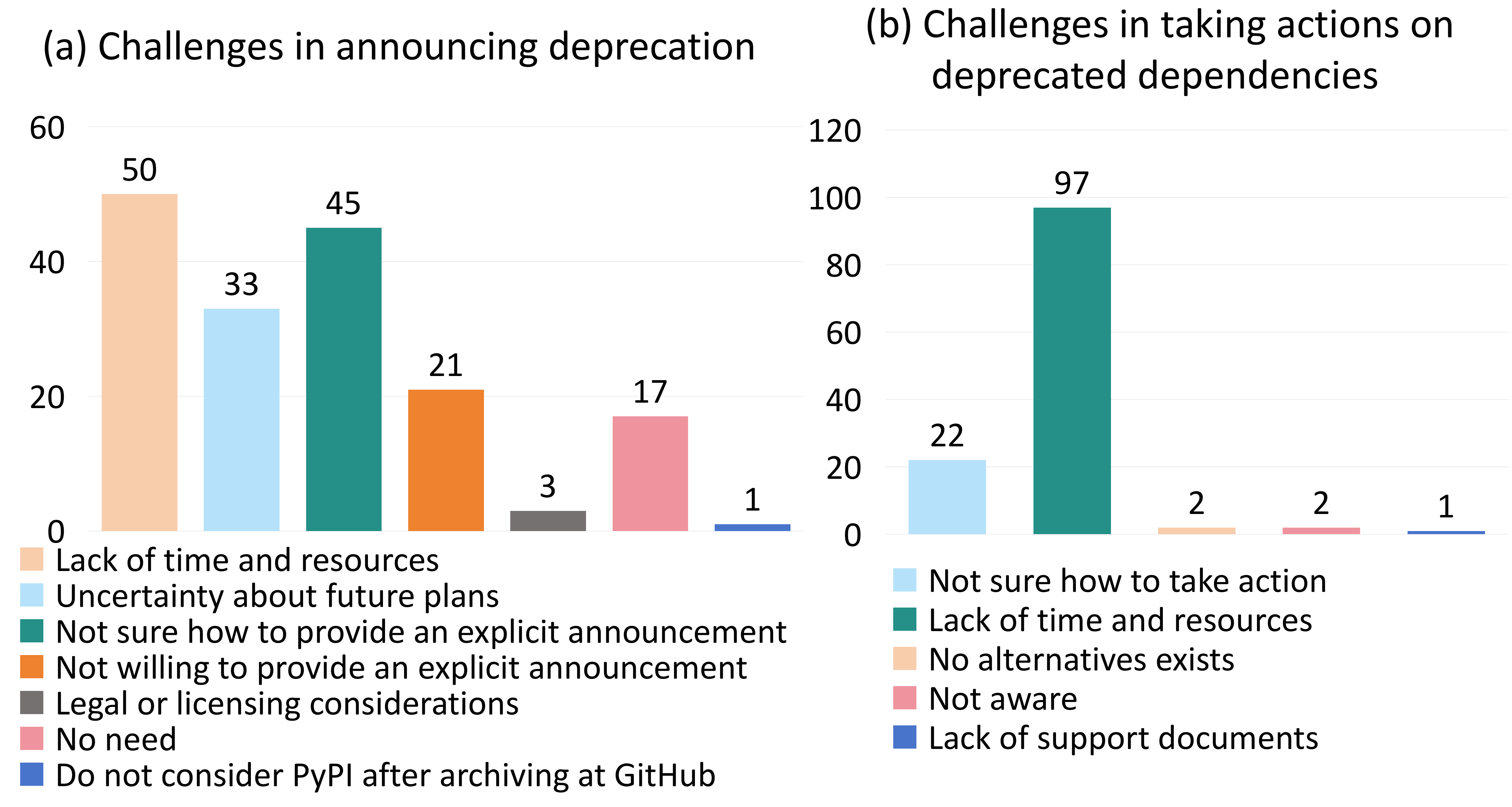}
    \caption{Results of Q5 in the Survey for inactive package developers and Q10 for in the Survey for deprecated package users}
    \label{fig:q5s1_q10s2}
\end{figure}
In conclusion, we have quantitatively and qualitatively verified the benefits of making a deprecation announcement, underscoring the necessity of such announcements.

\def\ChaLen{2.67cm}
\def\CoefLen{13cm}
\begin{table}
  \centering
  \caption{Summaries of linear mixed-effect model regressions  
  }
  \label{tab:results}
  \begin{tabular}{llll}
    \toprule
    \multicolumn{2}{l}{} &\textbf{Gain of Downstream}&\textbf{Gain of Unresolved}\\
    \multicolumn{2}{l}{} &\textbf{Dependencies} &\textbf{Issues (log)}\\
    \multicolumn{2}{l}{} & Model I & Model II  \\
    \cmidrule(lr){3-4}
    \multicolumn{2}{p{\ChaLen}}{\textbf{Characteristics}} &  & \\
    \multicolumn{2}{p{\ChaLen}}{Is deprecated} & -1.15 (0.34) ***   &  -0.17 (0.05) ***\\
    \multicolumn{2}{p{\ChaLen}}{Project stars (log)} &\ 0.00 (0.09) & \ 0.23 (0.01) ***\\
    \multicolumn{2}{p{\ChaLen}}{Project age (log)} &\ 0.07 (0.12)  & -0.03 (0.01)\\
    \multicolumn{2}{p{\ChaLen}}{Num. of releases (log)} & -0.05 (0.15)& -0.04 (0.02)\\
    \multicolumn{2}{p{\ChaLen}}{Num. of downstream dependencies ever (log)} & -0.82 (0.17) *** & \ 0.15 (0.03) ***\\
    \multicolumn{2}{p{\ChaLen}}{Num. of contributors (log)} & -0.68 (0.21) **& \ 0.01 (0.03) \\
    \multicolumn{2}{p{\ChaLen}}{Duration since last commit (log)} & -0.46 (0.12) * & \ 0.01 (0.03) \\
    \midrule
    \multicolumn{2}{p{\ChaLen}}{Num. of observation} &\ 1,365 & \ 955 \\
    \bottomrule
    Note: *$p<0.05$; **$p<0.01$; ***$p<0.001$
  \end{tabular}
\end{table}

\begin{tcolorbox}[boxsep=1pt,left=2pt,right=2pt,top=3pt,bottom=2pt,width=\linewidth,colback=white!90!black,boxrule=0pt, colbacktitle=white!,toptitle=2pt,bottomtitle=1pt,opacitybacktitle=0]
\textbf{Finding 4.} \textit{Deprecation announcements deter both new and existing users from adopting the packages. They also help minimize the occurrence of unresolved issues that may arise after the packages become inactive. Furthermore, they simplify the decision-making process for users when determining whether to adopt or remove the packages.}
\end{tcolorbox}



\subsection{RQ3: Why do inactive packages rarely release a deprecation announcement?}
\label{sec:effective}
Despite the benefits of having a deprecation announcement, the results of RQ1 indicate that most developers are unwilling to make such announcements.
Consequently, we further investigated the challenges hindering them from providing deprecation announcements in Q5 of Table~\ref{tab:questionnaire1}.
As Figure~\ref{fig:q5s1_q10s2}(a) illustrates, the most frequently mentioned challenge is the lack of time and resources.
Notably, 38.1\% of respondents expressed uncertainty about how to provide explicit deprecation announcements.
Interestingly, these two challenges are also commonly mentioned challenges by users when it comes to taking action on deprecated dependencies,
as shown in Figure~\ref{fig:q5s1_q10s2}(b).
These findings highlight the necessity of developing an official deprecation mechanism that reduces effort and provides explicit guidelines for both package developers and users.
Additionally, 28.0\% of respondents expressed uncertainty about future plans.
We also observed in the last open-ended question that some participants mentioned a lack of usage information or feedback about the packages, which may weaken their motivation to announce deprecation.
\begin{tcolorbox}[boxsep=1pt,left=2pt,right=2pt,top=3pt,bottom=2pt,width=\linewidth,colback=white!90!black,boxrule=0pt, colbacktitle=white!,toptitle=2pt,bottomtitle=1pt,opacitybacktitle=0]
\textbf{Finding 5.} \textit{Inactive package developers are unwilling to announce deprecation primarily due to the lack of time and resources, and uncertainty about how to make the announcement. In addition, they are unsure about the necessity of the deprecation announcement due to the lack of feedback.}
\end{tcolorbox}

\subsection{RQ4: What are the expectations of package developers and users regarding the future deprecation pattern?}
\begin{figure}[h]
    \centering
    \includegraphics[width=0.5\textwidth]{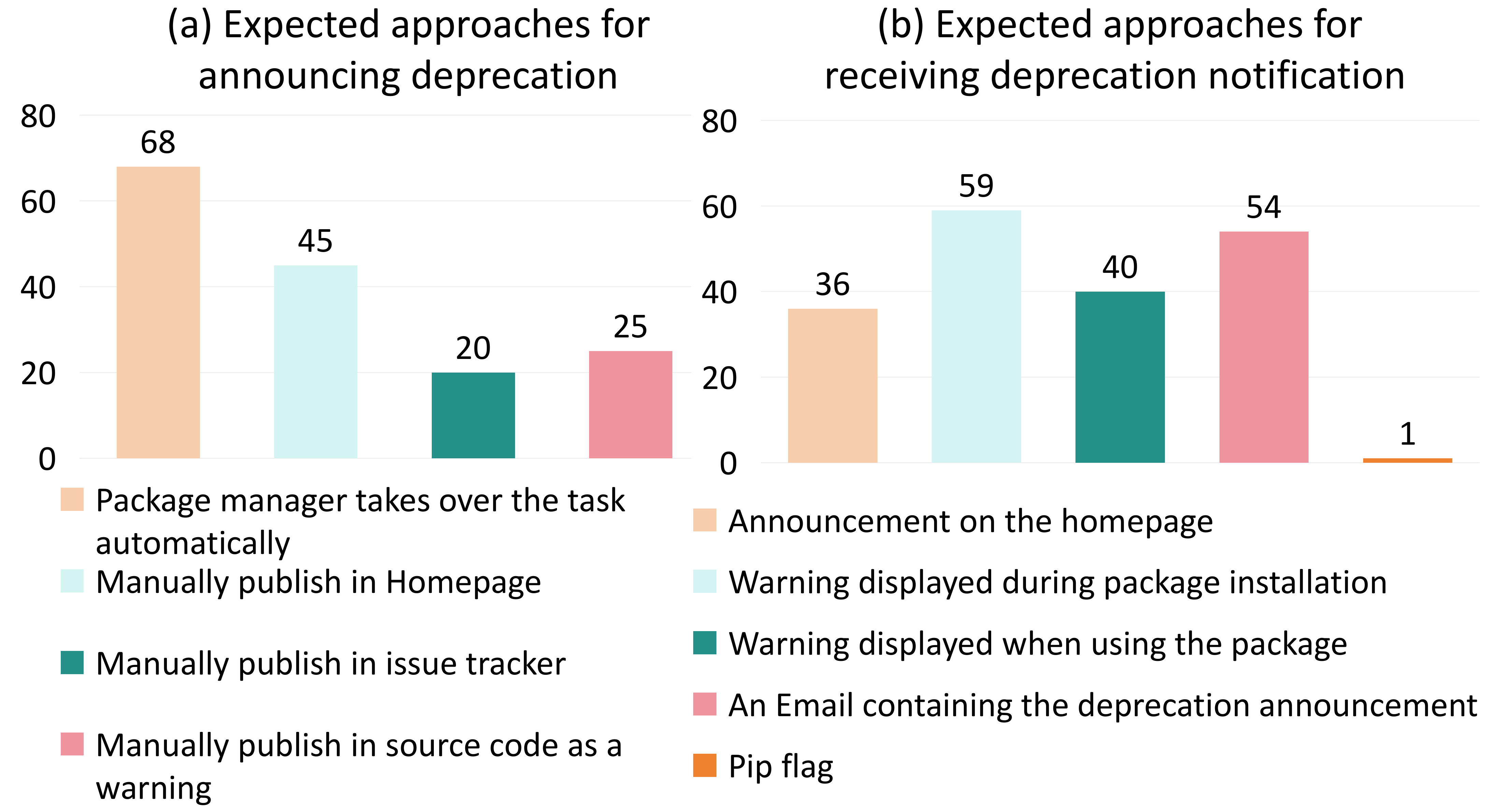}
    \caption{Results of Q6 in the Survey for inactive package developers and Q5 for in the Survey for deprecated package users}
    \label{fig:q6s1_q5s2}
\end{figure}

Now we present the results of Q6 from the survey for inactive package developers, which reveals their preferred method for announcing deprecation.
As Figure~\ref{fig:q6s1_q5s2}(a) depicts,
more than half of the developers expect the package manager to take over the tasks upon request, while there are still more than one-third of developers who will publish the deprecation announcement on the homepage manually.
Moreover, approximately 16.9\% are willing to manually announce in the issue tracker.
One developer stated that after announcing deprecation via GitHub, he no longer cares about the PyPI. This is understandable since GitHub is a powerful tool for developers, while PyPI serves as an archive. Therefore, developers naturally pay more attention to GitHub.
However, users may have a different perspective, resulting in less attention being paid to the corresponding GitHub repository compared to developers.
This emphasizes the necessity for an automated tool or feature that actively monitors the deprecation status on the corresponding GitHub repository.


Now moving to the perspective of deprecated package users,
the result in Section~\ref{sec:receiving_depre} shows that most users want to be notified about the deprecation.
Therefore, we further investigated the notification approaches they preferred in Q5 and the desired content for deprecation announcements in Q6.
Figure~\ref{fig:q6s1_q5s2}(b) depicts the most favored approaches: warnings displayed during installation, email notifications, and warnings displayed when using the package. 
All these methods eliminate the need for users to actively check for deprecation on their own.
Regarding the content to be included in the deprecation announcement, we found that most users want alternative solutions, and more than half of them want the reasons for depreciation.
However, it is rare for developers to know alternative solutions and thus they do not provide them in deprecation announcements.
Furthermore, one-third of users require a severity report on the deprecation, and one user also wants to know whether the package can be handed over to other maintainers.
In terms of additional support, apart from requested migration guidelines, over one-third of users also want guidance on taking over the package.
This becomes particularly effective when there are no potential alternatives, or when the migration process is prohibitively costly.
\begin{tcolorbox}[boxsep=1pt,left=2pt,right=2pt,top=3pt,bottom=2pt,width=\linewidth,colback=white!90!black,boxrule=0pt, colbacktitle=white!,toptitle=2pt,bottomtitle=1pt,opacitybacktitle=0]
\textbf{Finding 6.} \textit{Over half of the developers would like the package manager to handle deprecation announcements, but many still like to manually announce deprecation. As for users, they prefer to receive deprecation notifications through email and warnings, with alternative solutions and reasons included. They also require a severity report, guidelines for taking over the package, and migration guidance.}
\end{tcolorbox}

\section{Discussion and Implications}
In this section, we will discuss the observations and implications derived from the findings of our research questions, which can provide valuable insights for the community.

\textbf{A deprecation mechanism in the Python ecosystem that is developer-friendly and effective for users is needed.}
In Section~\ref{sec:estimating_impact}, we have demonstrated the benefits of having a deprecation announcement. Considering the low proportion of deprecation announcements and the lack of information, it is recommended to develop a deprecation mechanism that assists developers in effortlessly providing detailed deprecation announcements and effectively communicates the message to users.
Moreover, many developers have already announced or prefer to announce deprecation on GitHub, as it is a developer-centric tool and the platform where developers primarily focus their attention.
Therefore, it is also recommended for the future deprecation mechanism to establish a connection with code hosting platforms such as GitHub, which enables comprehensive monitoring of maintenance status and a more comprehensive collection of deprecation announcements.

\textbf{When no explicit deprecation announcement is available, having finer-grained categories of statuses that indicate different levels of risks may be beneficial.}
Figure~\ref{fig:q5s1_q10s2}(a) demonstrates that developers may choose not to provide a deprecation announcement for various reasons even if their packages have been without maintenance for a significant period and some are unlikely to be revived in the future.
While some packages may still function well, others may encounter compatibility or security issues.
These diverse statuses indicate varying levels of usability and security risks, suggesting that deprecation status can extend beyond binary options to include finer-grained categories.
If the risks associated with packages in different statuses can be evaluated and separated using defined boundaries, along with explicit descriptions, it would also assist package users in determining whether to adopt or remove a package even in the absence of explicit messages from developers. 
We recommend that further attention be given to this research topic.

\textbf{A more user-friendly practice is needed for users who want to take over the package.}
Although most users express a desire for alternative solutions to be included in the deprecation announcements, the findings in Section~\ref{sec:how_deprecations_been_made} reveal that only a low proportion of deprecation announcements actually provide such alternative solutions.
While previous works have focused on recommending alternative solutions at the package level and API level~\cite{he2021migrationadvisor,xi2019migrating}, one concern is the lack of potential alternative solutions.
In such situations, users can only maintain the dependencies themselves if they want to keep risks under control.
Instead of creating a new forked version, inheriting the deprecated package can retain users, saving them the effort of finding an alternative solution. Additionally, developers will not need to find new users since existing users can act as good testers of the package, reducing maintenance efforts.
However, a user who has decided to maintain the dependency himself mentioned that the requirements set by PEP~541~\cite{PEP541}, which is the official practice for inheriting the package, are too high. 
Furthermore, even if all the conditions are met, the review process by the Python Software Foundation official member may take a long time due to the volunteer/staffing shortage~\cite{PEP541_discussion}.
While PEP~541 can ensure security through its conservative strategies, it is suggested to have a supplementary mechanism with looser conditions to facilitate the takeover of packages. This mechanism should also inform users about the associated risks.

\textbf{Exploring better solutions for backward compatibility in the Python ecosystem is necessary.}
Despite the developers having completed the development of all the planned features, keeping the packages up-to-date remains a non-trivial task.
Some developers, as expressed in our surveys, have indicated that maintaining the package is labor-intensive and sometimes annoying.
It often requires more time than the initial development process itself.
In general, to prevent functionality and security issues, packages need to update their dependencies with patches. However, the frequent occurrence of breaking changes in these dependencies further complicates the maintenance process, leading to developer frustration.
From the perspective of upstream dependencies, backporting patches to older versions that have no breaking changes is also a labor-intensive task, especially considering the existence of multiple previous version tracks.
Therefore, it is recommended to explore the possibility of automatically backporting patches without causing breaking changes. By improving the backward compatibility of the Python ecosystem, maintenance tasks could become more manageable and potentially even automated.

\section{Related Work}

\subsection{Open-Source Sustainability} 
Early research focuses on attracting new contributors and retaining volunteers~\cite{guizani2022attracting,calefato2022will,fagerholm2014role}. 
However, abandonment is inevitable in the ecosystem. While some OSS projects have long lifespans (e.g., Linux, Eclipse, and Apache), others may become inactive or abandoned. 
Subsequent studies analyzed the nature of abandonment.
Khondhu et al.~\cite{khondhu2013all} and Ait et al.~\cite{ait2022empirical} found a significant population of inactive OSS projects on platforms like Sourceforge and GitHub. 
Further research delved into the reasons why developers give up on OSS projects~\cite{coelho2017modern,avelino2019abandonment,miller2019people}.
In addition, Valiev et al.~\cite{valiev2018ecosystem} investigated ecosystem-level factors that affect the sustainability of OSS projects.
Regarding the projects that were abandoned, Miller et al.~\cite{miller2023we} summarized the impacts and practices adopted to manage these projects via interviewing their users.
In this paper, we explore the demands of both package developers and users, providing insights into managing deprecation at the ecosystem level.



\subsection{Deprecation Mechanism} 
Early research focused on API-level deprecation mechanisms.
Sawant et al.\cite{sawant2018understanding} explored developers' needs regarding API-level deprecation mechanisms. However, subsequent findings revealed that API-level deprecation strategies in the Python ecosystem are often ad-hoc, resulting in deprecated APIs not being adequately handled \cite{wang2020exploring}.
For package-level and release-level deprecation, Cogo et al. and Li et al. have previously studied the deprecation mechanisms in the npm and Cargo ecosystems, respectively, and found that the existing mechanisms are rudimentary and in need of improvement \cite{cogo2021deprecation, li2022empirical}.
However, little is known about the status quo of package-level deprecation in the Python ecosystem.
This paper fills this gap and offers practical implications for establishing a deprecation mechanism within the Python ecosystem.

\section{Threats to Validity}

\subsection{Internal Validity}
We did not include some other practices that may be adopted to announce deprecation in this study,
such as the API-level deprecation mechanism~\cite{PEP702} and the yanking mechanism~\cite{PEP592}. 
This is mainly because these methods are not typically intended for deprecating an entire package, which limits their effectiveness for package-level deprecation.
For example, we scanned the metadata and identified 562 packages that yanked all their releases to deprecate themselves, and most of these (75\%) had only one release in total.
This suggests that such approach is less user-friendly and less recognized than other unofficial practices (e.g., publishing deprecation announcements on GitHub README or issues), especially for packages with multiple releases.
Consequently, we did not consider PEP 702 and the yanking mechanism in our study.

When conducting the study, we encountered challenges in collecting comprehensive data from the sources. One challenge involved collecting dependency information for the entire Python ecosystem, which can be difficult. Various methods for obtaining dependency information have been adopted in many papers. A common data source used by researchers is the data dump from Libraries.io \cite{libraries2020data}. However, this data dump suffers from certain weaknesses. Firstly, it is outdated, with its last dump released on January 12, 2020, lacking information on packages and releases in recent years. Additionally, it may contain missing dependencies, a common and easily detectable issue.
Researchers can also access dependency information by utilizing the API provided by PyPI \cite{PyPI2023api}. However, this source also suffers from incompleteness, similar to Libraries.io. To overcome these limitations, we leveraged the code provided by Valiev et al.~\cite{valiev2018ecosystem} to extract dependency information from distribution files on PyPI. This approach surpasses the previous methods in terms of timeliness and completeness. 
Additionally, we encountered rare instances of missing distribution files on PyPI and missing issues on GitHub due to deletion. However, such occurrences are not substantial enough to affect the validity of our results.
When identifying deprecated packages from inactive ones, we adopt a keyword search strategy to identify deprecation announcements from related README and issues. 
However, this strategy can miss some packages that did have a deprecation announcement without those keywords and incorrectly consider these packages as inactive ones.
To validate this risk, we sampled 383 inactive packages without keywords. The sample size was determined using the sample size calculator, following the same procedure described in Section~\ref{sec:effect}. Subsequently, two of our authors checked for any deprecation announcements that might have been missed by our previous identification method. We found that there are 6 packages where this was the case, indicating that the accuracy of our previous method is 98.4\% with a 95\% confidence interval.

The two surveys are also influenced by social desirability bias~\cite{nederhof1985methods} and self-selection bias~\cite{smart1966subject}. To alleviate these biases, we tried to collect more responses by sending invitation emails to people who may be interested in the study. In addition, we informed the respondents that their responses would remain anonymous.

\subsection{External Validity}
Our study focuses on analyzing the current deprecation pattern in the Python ecosystem. It is important to note that since there is currently no existing deprecation mechanism for this ecosystem, the results obtained may not necessarily generalize to other programming language ecosystems that do have a deprecation mechanism in place.
For this work, we specifically concentrated on packages that have corresponding GitHub repositories. While it is true that some packages may be hosted on other code control platforms such as GitLab, most of the repository-related information we collected, such as issues and stars, is also available on other platforms.
\section{Conclusion}
In this paper, we have presented an analysis of package-level deprecation in the Python ecosystem, providing insights into its status quo.
We further demonstrated the impact of making deprecation declarations.
To offer valuable guidelines for the future development of package-level deprecation mechanisms, we have conducted an investigation into the challenges that package developers and users faced, as well as expectations on the future deprecation pattern. 
Additionally, we have released a new dataset, serving as a foundation resource for future deprecation mechanisms.
An in-depth discussion has been carried out to serve as a guideline reference for future research and the advancement of package-level deprecation mechanisms.


\section{Data Availability}
The dataset we collected and the related source code can be found at \url{https://doi.org/10.5281/zenodo.13335360}.
\section{Acknowledgment}
This paper was supported by the National Natural Science Foundation of China (No. 62102340), Guangdong Basic and Applied Basic Research Foundation (No. 2024A1515010145), and Shenzhen Research Institute of Big Data Innovation Fund (No. SIF20240009).
\balance

\bibliographystyle{IEEEtran}
\bibliography{IEEEabrv,Reference}
\end{document}